# Tailoring Quasi-Bound States in the Continuum for Infrared Photodetection in Black Phosphorus

Xiao Liu[1,2 †], Tianxiang Zhao[3 †], Ting Wang[2,4], Junsheng Xu[4,5], Junyong Wang[5], Kai Zhang[5], Hongliang Li[2*], Xuechao Yu[2,4*], and Junjia Wang[1,3*]

[1] School of Integrated Circuits, Southeast University, Wuxi 214026, P. R. China;

[2] Key Laboratory of Multifunctional Nanomaterials and Smart Systems, Suzhou Institute of Nano-Tech and Nano-Bionics, Chinese Academy of Sciences, Suzhou 215123, P. R. China;

[3] School of Electronic Science and Engineering, National Research Center for Optical Sensing/Communications Integrated Networking, Southeast University, Nanjing 210096, P. R. China;

[4] Nano Science and Technology Institute, University of Science and Technology of China, Hefei 230026, P. R. China;

[5] i-Lab, Suzhou Institute of Nano-Tech and Nano-Bionics, Chinese Academy of Sciences, Suzhou 215123, P. R. China;

[†] These authors contributed equally to this work. [*] Corresponding authors.

**ABSTRACT**

High-performance infrared photodetection underpins various applications spanning surveillance, environmental monitoring, optical communications and biomedical imaging. However, conventional bulk detectors remain limited by poor spectral tunability, mechanical rigidity, and high dark currents, motivating the pursuit of low-dimensional material platforms such as graphene and transition metal dichalgenides. Black phosphorus (BP) is particularly compelling in this context, owing to its thickness-tunable direct bandgap, high carrier mobility, and pronounced in-plane anisotropy. Nevertheless, its atomically thin nature inherently restricts light absorption, posing a fundamental bottleneck for device performance. Here, we demonstrate quasi-bound states in the continuum (quasi-BICs) within a dielectric metasurface integrated with BP, enabling strongly enhanced and spectrally selective light-matter interactions. By introducing controlled symmetry breaking at the unit-cell level, high-quality-factor resonances are realized, resulting in pronounced electromagnetic field confinement within the BP layer. This resonant enhancement substantially increases photocarrier generation while preserving the intrinsic polarization anisotropy of BP, which elucidates a robust pathway for overcoming the optical absorption bottleneck in anisotropic 2D optoelectronics via quasi-BIC platforms.

# 1. Introduction

Infrared photodetectors play a pivotal role in contemporary optoelectronic systems, enabling advancements in telecommunications, imaging, sensing, and quantum technologies.[1-4] However, conventional bulk photodetectors are increasingly constrained by limited spectral tunability, mechanical rigidity, and high power consumption, particularly as device miniaturization and performance demands continue to intensify. To overcome these limitations, low-dimensional materials have been actively explored for next-generation photodetection, including graphene, black phosphorus (BP), transition metal dichalcogenides (TMDs), and MXene.[5-9] Among these two-dimensional materials, BP is distinguished by virtue of its unique combination of exceptional properties, most notably a layer-tunable direct bandgap spanning a broad spectral range from visible to infrared, coupled with high charge carrier mobility.[10-19] Furthermore, its puckered crystal lattice gives rise to pronounced in-plane anisotropy, resulting in strongly polarization-dependent electrical and optical responses along the armchair and zigzag crystallographic orientations.[20] Nevertheless, the atomically thin nature of BP inherently limits light absorption and quantum efficiency, further exacerbated by rapid carrier recombination. These factors collectively impose a fundamental bottleneck on device performance, necessitating effective strategies to enhance light-matter interactions without compromising intrinsic material properties.

Metasurfaces have emerged as a powerful platform to overcome these limitations by manipulating light at subwavelength scales, enabling enhanced optical absorption and tailored light-matter interactions.[21-24] Comprising arrays of subwavelength nanostructures, metasurfaces enable precise control over the amplitude, phase, and polarization of light.[25-27] This powerful wavefront manipulation capability makes them a versatile platform for controlling light-matter interactions at the nanoscale, motivating their widespread use in boosting the performance of photodetectors, especially integrated with two-dimensional materials.[22,28,29] However, the optical performance and efficiency gains offered by conventional metasurfaces are frequently constrained by intrinsic optical losses, caused by radiation losses due to the coupling between resonant modes and free space. Various design and material strategies have been proposed to mitigate this limitation, including high refractive index dielectric material, van der Waals heterojunction, and phase-change material $VO_2$.[30-34] Bound states in the continuum (BICs), the non-radiating resonant modes endowed with a theoretically infinite quality factor, provide exceptional capabilities for enhancing light-matter interaction and suppressing radiative loss, rendering them highly attractive for integration with metasurfaces to achieve unprecedented optical control.[35] By introducing controlled symmetry breaking or tuning geometric parameters, symmetry-protected BICs can be converted into leaky resonances (quasi-BICs), enabling controlled radiation leakage while sustaining high-quality factors. This results in strongly localized electromagnetic fields that markedly promote light–matter interactions.[36-39] Owing to their ability to confine light and dramatically strengthen light-matter interactions, quasi-BICs have sparked a surge of interest across diverse fields, ranging from low-threshold lasing and ultrasensitive sensing to nonlinear optics.[40-43] More recently, quasi-BIC-based metasurfaces have been explored to augment the performance of photodetectors based on transition metal dichalcogenides (TMDs).[44,45]

In this work, we demonstrate a metasurface-integrated BP photodetector driven by quasi-BICs that amplifies photoresponsivity by enhancing light absorption and suppressing radiative losses at target wavelengths. Guided by numerical simulations, we engineered the quasi-BIC resonance to align with the telecommunication wavelength of 1550 nm. The metasurface integration yields significant enhancements in photocurrent and responsivity at the resonant wavelength relative to pristine BP. These findings indicate that the metasurface-driven approach enhances light absorption while preserving BP's intrinsic polarization characteristics, establishing a route toward high-efficiency, polarization-resolved photodetectors for next-generation applications.

## 2. Results and discussions

### 2.1. Design and characterization of the BP photodetector

The device comprises a silicon-on-insulator (SOI) substrate, a lithographically defined metasurface, a BP flake capped with hexagonal boron nitride (h-BN), and Cr/Au contact electrodes fabricated on both sides of the metasurface, as shown in **Figure 1(a)**. The metasurface composed of asymmetric nanorod is patterned directly in the top layer of the SOI substrate (500 nm top silicon layer and 2 μm buried $SiO_2$ layer). A BP flake is transferred onto the metasurface and encapsulated with an h-BN layer, thereby effectively passivating the flake against ambient oxidation. Under optical illumination, photocarrier generation within the BP layer is driven by the strongly localized electromagnetic field associated with the quasi-BIC resonance supported by the underlying metasurface. Photodetection is achieved by applying a bias voltage across the electrodes and collecting the resulting photocurrent. Fabrication details are provided in the **Materials and methods** and **Figure S1**.

**Figure 1(b)** depicts the crystallographic structure of BP, highlighting the characteristic layered puckered lattice, is associated with high carrier mobility and pronounced anisotropy.[10,46] The quasi-BIC resonance is induced by breaking the in-plane symmetry of the unit cell, which is achieved via the introducing an arc-shaped notch into the silicon nanopillar and precisely controlling its depth, as depicted in **Figure 1(c)**.[35,44,47] Through numerical simulation-based optimization of the structural geometry, the optimized asymmetry parameter was determined to be $d = 60$ nm, followed by experimental verification. As illustrated in the optical image and SEM images (see **Supporting Information**, **Figure S2**), arrays of Si nanorod pairs were fabricated with systematically varied asymmetry ($d$) from 40 to 140 nm in increments of 20 nm. The enlarged SEM images of the fabricated samples with $d = 60$ nm in **Figure 2(e)** demonstrate strong agreement with the corresponding numerical simulation, confirming the high fidelity of the nanofabrication process. Finally, by balancing the trade-offs between the Q-factor, coupling efficiency, and fabrication tolerance, unless otherwise specified, the geometric parameters of the metasurface were fixed as follows: $x = 1000$ nm, $y = 470$ nm, $z = 500$ nm, $w = 115$ nm, $d = 60$ nm. **Figure 1(d)** displays the optical microscopy image of the device morphology, and **Figure 1(e)** exhibits the AFM height profile extracted along the black dashed line in **Figure 1(d),** confirming a BP thickness of approximately 12 nm. Raman spectroscopy performed in the 300-500 cm$^{-1}$ range reveals distinct peaks corresponding to BP, as depicted in **Figure 1(f)**. The spectrum exhibits three distinct Raman characteristic peaks at 361, 438, and 466 cm$^{-1}$, corresponding to the $A_{1g}$, $B_{2g}$, and $A_{2g}$ vibration modes in BP, respectively. These peaks are in good agreement with reported values for pristine BP, and their sharp linewidths indicate high crystalline quality of the sample.[9,20]

Based on the above characterizations, the introduced asymmetry in the metasurface excited a quasi-bound state in the continuum (quasi-BIC), leading to enhanced local field. Thereby, the optical absorption of black phosphorus was strengthened, and the photodetection proceeds through three sequential processes: (i) excitation of quasi-BIC modes near 1550 nm, which tightly confine and strongly enhance the optical fields in the BP layer; (ii) enhanced light absorption in BP, producing a high photocarrier density; and (iii) separation of the photogenerated carriers by an applied bias, followed by their transport toward the electrodes to generate photocurrent. Controlled geometric asymmetry ($d > 0$) breaks the in-plane symmetry, opening a radiative channel that converts a symmetry-protected BIC into a high-Q quasi-BIC resonance. By combining resonant field intensification, efficient photocarrier generation, and high-quality factors, the asymmetric metasurface enables a route toward highly responsive and spectrally selective sub-bandgap photodetection in multilayer BP.

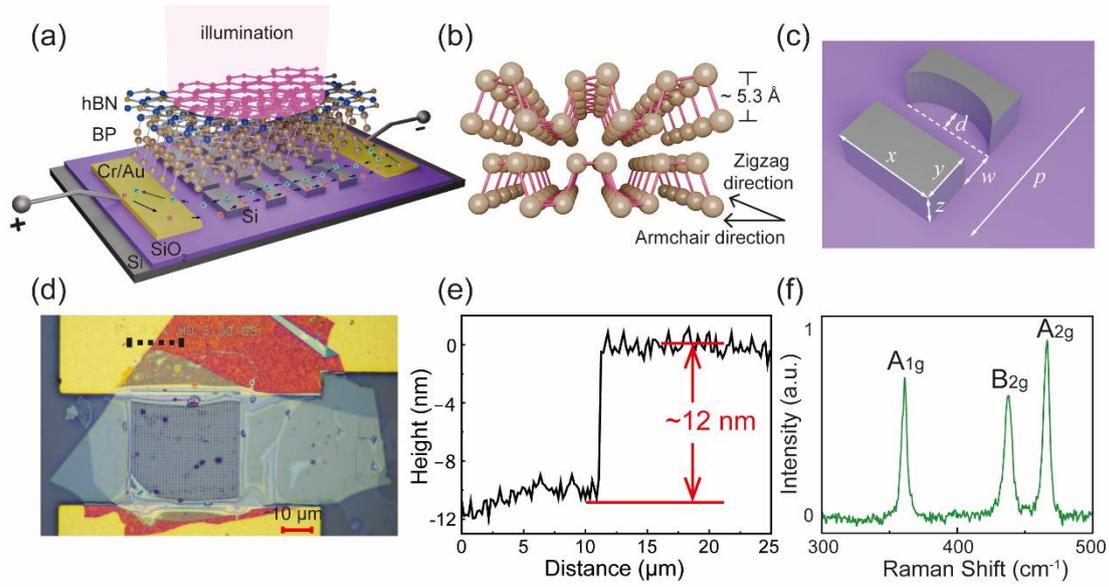

**Figure 1. Architecture of the multilayer BP photodetector enabled by quasi-BIC. (a)** Schematic diagram of the photodetector based on a hybrid structure integrating a quasi-BIC metasurface with multilayer BP. **(b)** Perspective view of the BP crystal structure. The interlayer spacing is approximately 5.3 Å. **(c)** Key geometrical parameters of the metasurface design, including the length ($x$), width ($y$), height ($z$) and spacing ($w$) of the array units, as well as the width ($d$) of the arc-shaped notch. **(d)** Optical micrograph of the fabricated device. The top h-BN layer is patterned to encapsulate the BP, protecting the air-sensitive material from degradation and defining the active device area. **(e)** AFM height profile of the multilayer BP measured along the black dashed line in the **1(d)**. **(f)** Normalized Raman spectrum of the multilayer BP in the 300–500 cm$^{-1}$ range.

**2.2. Simulations and characterizations of the quasi-BIC**

To further elucidate the underlying mechanism of the quasi-BIC effect introduced by the asymmetric metasurface, numerical simulations and characterization of the quasi-BIC were conducted. The metasurface design was guided by a series of optical and electromagnetic simulations, which established the optimal structural parameters and provided a simulation-based foundation for the subsequent device fabrication process. The metasurface was then fabricated and characterized to confirm that the introduced asymmetry gives rise to the intended quasi-BIC mode. To elucidate the modal characteristics, the photonic band structure was computed across three-dimensional momentum space, as shown in **Figure 2a**. The dispersion surface features a smooth band-edge extremum at a normalized frequency of ω ≈ 0.65. Near the Γ point, the small band curvature indicates a suppressed group velocity, yielding a slow-light effect that promotes the local density of states and strengthens light-matter interaction. The quasi-BIC establishes an ideal photonic platform featuring an ultrahigh quality factor, pronounced dispersion, and exquisitely tunable radiative decay. The concomitant slow-light effect thus emerges as a quintessential manifestation of these attributes, directly reflected in the reduction of the group velocity.[48,49] Consequently, the engineered band structure near the Γ point, caused by combining the slow-light effect with the quasi-BIC resonance, provides a promising route to increase optical absorption in BP and optimize device performance.

**Figure 2(b)** shows the far-field polarization distribution in the $k_x$-$k_y$ plane for the asymmetric structure ($d$ = 60 nm), illustrating the topological origin of the high-Q resonance. Despite the broken in-plane symmetry, the polarization vectors retain a vortex texture centered at the Γ point, demonstrating the robustness of the mode. While the asymmetry opens a leaky radiation channel, the mode fundamentally preserves its bound-state nature, thereby maintaining a high-quality factor. The Q-factor heat map in **Figure 2(c)** confirms that Q peaks above $10^6$ at the center and decreases toward the periphery, which enables robust confinement of the local electromagnetic field, synergistically amplifying the optical absorption efficiency within the BP. The coexistence of the polarization vortex and the high, yet finite, Q indicates operation in the quasi-BIC regime. Although ultra-high Q factors maximize local field confinement, they impose stringent fabrication requirements: small dimensional deviations can shift the resonance away from the target 1550 nm.

The chosen asymmetry parameter of $d = 60$ nm yields a quasi-BIC mode with a moderately high Q factor, achieving sufficient field confinement to enhance BP absorption while maintaining a spectral bandwidth that is tolerant to fabrication imperfections. The electric field distribution within the symmetry-broken unit cell was evaluated via simulations (**Figure 2(d)**). The fields are highly non-uniform, with strong confinement near the arc-shaped notch and negligible intensity in the periphery, demonstrating effective quasi-BIC field confinement for enhanced light-matter interaction. Following the simulation-guided design, the metasurface was fabricated and characterized to verify the formation of the quasi-BIC mode. **Figure 2(e)** presents the SEM images of the fabricated device array and magnified unit cells, allowing evaluation of the nanoscale architecture. The images reveal a highly uniform array of rectangular structures with precise dimensions. The measured length and width of the unit cell are approximately 930 nm and 430 nm, respectively, consistent with the design targets. Crucially, the asymmetry parameter, defined by the arc-shaped notch depth, is measured to be $d \approx 60$ nm, demonstrating the high fidelity and controllability of the fabrication process. After finalizing the design dimensions through simulation and confirming the fabricated structure via SEM, angle-resolved spectroscopy was performed to provide preliminary experimental evidence for the existence of the quasi-BIC mode in the metasurface, mapping its reflection intensity as a function of incident angle and wavelength, as shown in **Figure 2(f)**. Under normal incidence, a sharp, high-intensity resonance peak is observed near 1550 nm, corresponding to the excitation of the quasi-BIC mode. As the incident angle deviates from normal, the resonance wavelength shifts noticeably, exhibiting a dispersive behavior, while the resonance intensity preserved within a narrow angular range. This strong angular dependence confirms the high-Q nature of the mode and highlights the stringent momentum-matching requirements for coupling. Additional angle-resolved spectroscopy results for different asymmetry parameters ($d$) can be seen in **Figure S3**. These findings demonstrate that the structure supports a high-quality-factor quasi-BIC mode near 1550 nm, highlighting its strong wavelength selectivity and pronounced angular sensitivity.

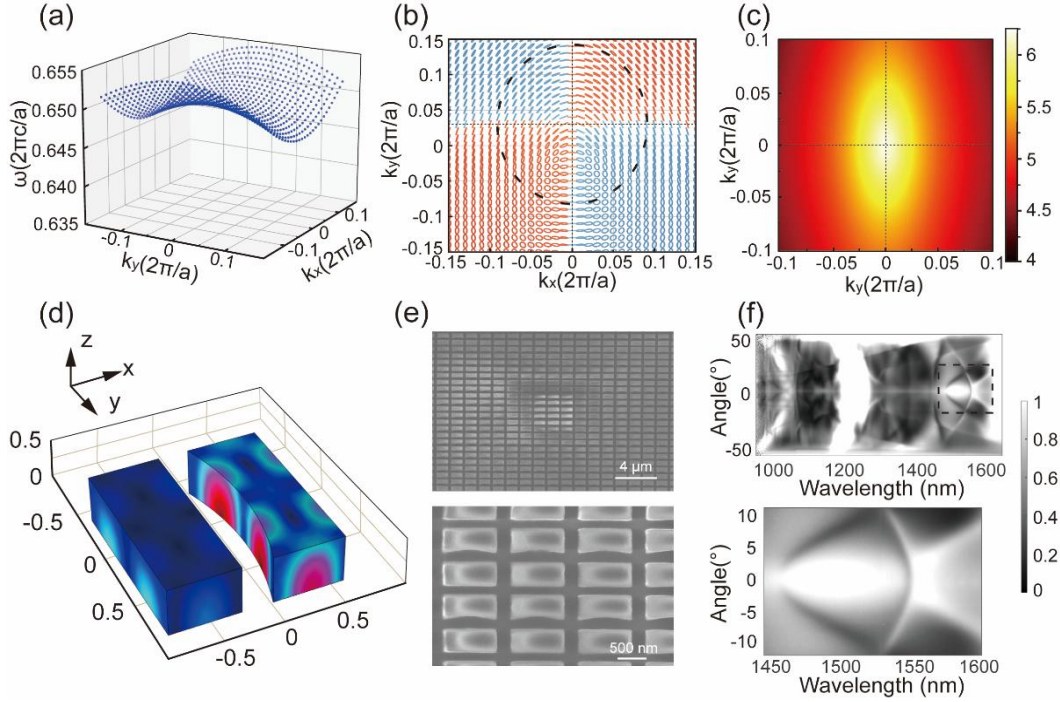

**Figure 2. Design simulation and experimental verification of the quasi-BIC metasurface. (a)** Simulated photonic band structure in three-dimensional momentum space. **(b)** Far-field polarization distribution in momentum space. The polarization vectors maintain a vortex texture converging at the momentum space center, indicating the topological robustness of the quasi-BIC mode. **(c)** Simulated Q distribution in the k-parameter space. **(d)** Simulated electric field intensity distribution within the symmetry-broken unit cell ($d = 60$ nm), highlighting strong field confinement near the arc-shaped notch. **(e)** SEM images of the fabricated metasurface array and a magnified view of unit cells with $d \approx 60$ nm. **(f)** Angle-resolved spectroscopy as a function of incident angle and wavelength. The bottom panel presents the enlarged view with wavelength from 1450 nm to 1600 nm, showing that the reflection intensity suddenly increases when the incident angle deviates from normal incidence or the wavelength deviates from 1550nm.

## 2.3. Characterization of the infrared BP photodetector

Guided by the simulation and optical characterization results, devices were fabricated by transferring black phosphorus onto the metasurface. A set of photodetectors was prepared, and the device corresponding to $d = 60$ nm was selected for detailed photoelectrical testing. The photoresponse of the photodetector was systematically measured to investigate and quantify the enhancement conferred by the quasi-BIC metasurface, especially by comparing devices with (w/m) and without (w/o) the metasurface. The optical system for detecting photoresponse is depicted in **Figure 3(a)**. To characterize the fabricated quasi-BIC-enhanced BP infrared photodetector, comparative photoresponse measurements were performed between the metasurface-integrated device and the pristine BP reference device. The time-dependent photocurrent characteristics were recorded under illumination at wavelengths of 808 nm, 1064 nm, 1342 nm, and 1550 nm, with the corresponding incident optical powers marked above each curve as depicted in **Figure 3(b)**. Additionally, the dark current-voltage characteristics under various bias conditions are provided in **Figure S4**. Our results demonstrate that the metasurface-integrated sample yields consistently higher photocurrent across all tested wavelengths compared to the pristine BP device, while maintaining temporal stability, with a pronounced enhancement specifically at the resonant wavelength of 1550 nm.

In photoelectrical measurements using a 50% duty cycle (light on/off), illumination triggers a rapid current increase, followed by a slower rise until the light is switched off, whereupon the current drops sharply. This behavior suggests the presence of two distinct physical processes: a fast, strong initial response and a slower, weaker subsequent component. The immediate current jump upon laser illumination stems from the prompt generation and separation of photocarriers. Photons with energies above the bandgap of BP are intrinsically absorbed, creating electron-hole pairs. Driven by the external bias, these photogenerated carriers drift rapidly across the channel. The high carrier mobility of BP ensures a short transit time, yielding an almost instantaneous photocurrent. Under continuous illumination, the current exhibits a gradual increase instead of saturation. This slower dynamic is primarily attributed to trap-state filling mechanism and the photothermal effect, with the former serving as the dominant part. Trap states, introduced by defects and impurities, initially capture a fraction of the photogenerated carriers, reducing the free-carrier density. As illumination continues, these trap states become gradually occupied, thereby reducing the capture rate and increasing the free-carrier concentration and conductivity. Additionally, local heating from the continuous laser irradiation can narrow the bandgap and increase the intrinsic carrier concentration through thermal excitation. Given the relatively low thermal conductivity of BP, the accumulated heat dissipates slowly, resulting in a gradual temperature rise that correlates with the slow photocurrent growth. These combined mechanisms account for the observed current dynamics during extended illumination. The photoresponsivity at 808 nm, 1064 nm, 1342 nm, and 1550 nm wavelengths, derived from current-time curves normalized to optical power, is summarized in **Figure 3(c)**. At 1550 nm, the responsivity reaches 106.85 µA/W with the metasurface, representing an approximately eight-fold enhancement over the pristine BP device. This improvement is attributed to the localized electromagnetic field amplification induced by the quasi-BIC resonance. To further analyse the device operation, the bias-dependent enhancement factor was investigated as plotted in **Figure 3(d).** The results reveal that the relative enhancement factor at 1550 nm peaks at 7.8 under a bias voltage of −0.5 V, whereas factors of 3.9, 2.4, and 2.3 are observed at −1 V, 0.5 V, and 1 V, respectively. This bias-dependent behavior reveals an intrinsic link between the operational mechanism of the metasurface and the carrier extraction dynamics. The maximum performance at −0.5 V can be attributed to the optimal band alignment and efficient separation of photogenerated carriers before the onset of substantial dark currents, whereas higher bias voltages introduce elevated background noise and potential scattering events that dilute the relative optical enhancement. Consequently, unless otherwise specified, an optimized bias voltage of −0.5 V was adopted for all subsequent photoelectrical characterizations.

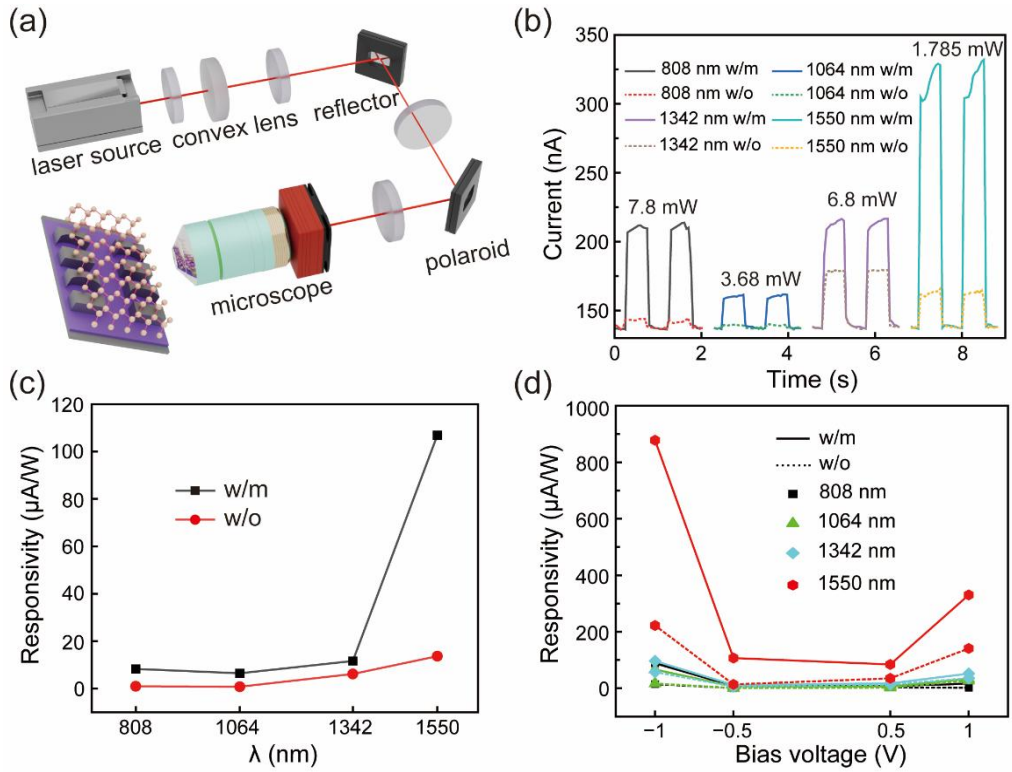

**Figure 3. Photoelectrical characterization of the BP photodetector.** **(a)** Schematic diagram of the optical setup for photoelectrical measurements. **(b)** Time-resolved photocurrent responses of the devices with and without the metasurface under continuous illumination at 808 nm, 1064 nm, 1342 nm, and 1550 nm. **(c)** Calculated photoresponsivity of the metasurface-integrated device compared to the pristine BP reference across various incident wavelengths. **(d)** Bias-dependent responsivity enhancement of the photodetector across different wavelengths. The enhancement factor at 1550 nm peaks at 7.82 under an optimal bias voltage of −0.5 V.

To examine the impact of optical power on the photocurrent response of a BP-based photodetector at 1550 nm, time-resolved photocurrent measurements were conducted across an optical intensity range of 0.75 to 2.5 (corresponding to optical powers ranging from 0.0003 mW to 1.785 mW), as shown in **Figure 4(a) and 4(b)**.

The experimental results demonstrate an approximately linear increase in photocurrent with rising optical power up to 1.395 mW (an optical intensity of 2). Beyond this threshold, the growth rate declines markedly, indicating the onset of photocurrent saturation. This saturation behavior suggests a transition in carrier recombination dynamics. Under high-power illumination, the photogenerated carrier concentration rises substantially, making Auger recombination the dominant loss mechanism.[50,51] Given that the Auger recombination rate scales with the cube of the carrier concentration, this strong nonlinear dependence leads to a drastic reduction in carrier lifetime ($\tau$) as the light intensity increases.[52,53] Consequently, although the carrier generation rate increases linearly, the steady-state carrier concentration exhibits sublinear growth and eventually saturates. Furthermore, photothermal effects constitute an additional contributing factor under intense illumination. Localized heating of the device intensifies lattice vibrations and phonon scattering, which reduce carrier mobility.[54,55] Additionally, the temperature elevation can modify the bandgap of the BP layer and increase the thermal excitation rate of traps and defects, further complicating the recombination dynamics.[56,57] These thermal mechanisms act synergistically with Auger recombination, promoting the saturation behavior observed at high optical power.

Controlling polarization response is a highly desirable capability in photodetectors, enabling critical applications ranging from secure communications to hyperspectral imaging.[58-60] Metasurfaces, serving as subwavelength-engineered interfaces, present a versatile means to tailor polarization functionalities in photonic devices.[61-63] To this end, polarization-resolved photocurrent measurements are indispensable for assessing whether the metasurface enhances the intrinsic anisotropy or introduces a new polarization-selective mechanism. In order to evaluate the polarization-dependent photoresponsivity at 1550 nm, photocurrent measurements were conducted across polarization angles ranging from 0° to 360°, comparing the metasurface-integrated device with the pristine BP reference as demonstrated in **Figure 4(c)**. Our findings demonstrate a pronounced polarization dependence in both cases, characterized by peak photocurrents at 170° and 350° and minima at 75° and 255°, along with a substantial enhancement in photocurrent magnitude when the metasurface is incorporated. The preserved angular correspondence of the photocurrent maxima and minima between the two configurations reveals that the polarization-selective mechanism remains critically anchored to the crystallographic axes of BP. BP exhibits intrinsically orthogonal armchair and zigzag crystallographic axes, as shown in **Figure 1(b)**, giving rise to pronounced in-plane anisotropy in both optical absorption coefficients and carrier mobilities under different polarization states of illumination. Furthermore, the polarization-sensitive behavior of BP is modulated by multiple factors, including the excitation wavelength, scattering configuration, and material layers.[64] The metasurface, which excites the quasi-BIC mode, possesses $C_S$ point group symmetry, characterized by mirror symmetry with respect to the *x*-axis, the alignment direction of the two constituent subunits within each unit cell. Consequently, the electromagnetic field distributions are mirror-symmetric about the *x*-axis, and all optoelectronic characterizations are conducted under normal incidence, corresponding to the Γ-point mode. Notably, the geometrical perturbation introduced to activate the quasi-BIC resonance is deliberately kept minimal; it merely opens a weak radiative channel without inducing appreciable additional structural anisotropy. Thus, the enhanced optoelectronic response above is attributed to intensified localized electromagnetic fields caused by quasi-BIC, rather than modifications in anisotropy. Therefore, the metasurface architecture amplifies the intrinsic polarization sensitivity of BP while preserving its original polarization orientation, rather than redefining the polarization axis, providing an effective strategy for realizing high-performance, polarization-sensitive photodetectors.

**Figure 4(d)** and **4(e)** exhibit the rise and fall times of the pristine BP reference and the metasurface-integrated device, respectively. Additionally, femtosecond transient absorption (TA) spectroscopy was conducted to elucidate the effect of the quasi-BIC metasurface on the carrier transient lifetime dynamics, as depicted in **Figure S5**. The intense field localization produced by the metasurface yields an exceptionally high photogenerated carrier density within a highly confined volume of the BP active layer. Consequently, these carriers have a higher probability of being captured by defect states, which are either intrinsic to the BP crystal or introduced during the metasurface fabrication process. Carriers trapped in deep level defects can be released slowly through thermal activation and tunneling, and undergo delayed recombination. As a result, the photocurrent exhibits a long decay tail, which directly translates into prolonged rise and fall times. While our design demonstrates enhanced photoresponsivity, this improvement comes at the expense of response speed, highlighting a fundamental performance trade-off commonly encountered in photodetector engineering.

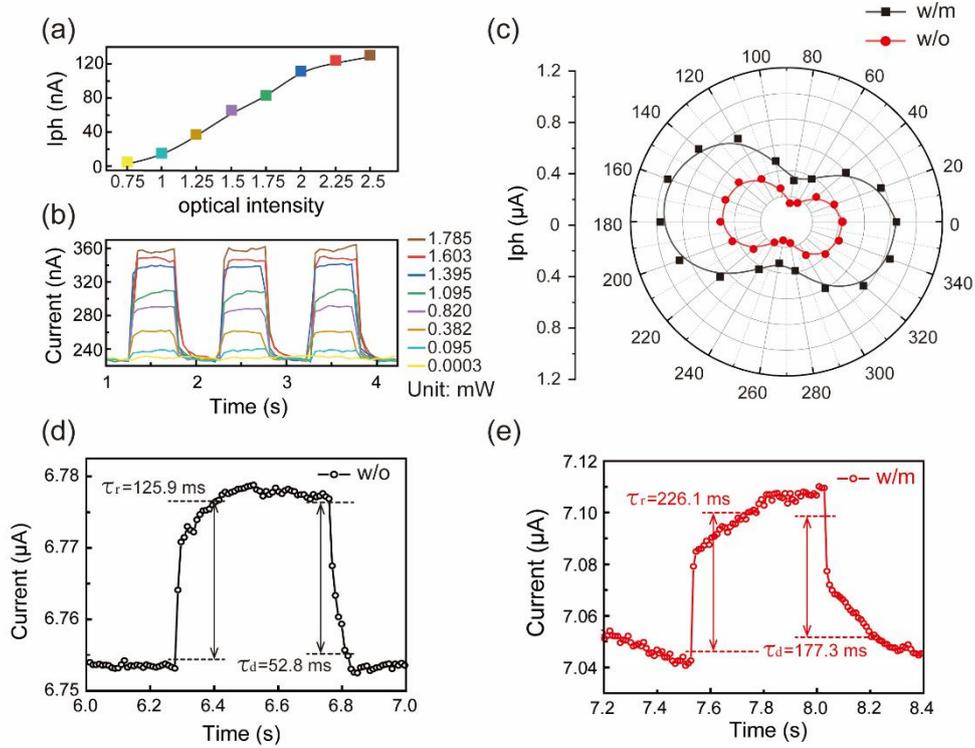

**Figure 4. Optical power dependence, polarization response, and temporal dynamics. (a)** Photocurrent as a function of incident optical intensity under 1550 nm illumination. **(b)** Time-resolved photocurrent responses at 1550 nm under various incident optical powers. **(c)** Polarization-resolved photocurrent of photodetector across polarization angles ranging from 0° to 360° under 1550 nm illumination. **(d)** Temporal response of the pristine BP reference device at a bias voltage of 0.1 V under 1550 nm illumination, exhibiting a rise time $\tau_r$ = 125.9 ms and a decay time $\tau_d$ = 52.8 ms. **(e)** Temporal response of the metasurface-integrated device at a bias voltage of 0.1 V under 1550 nm illumination, exhibiting a rise time $\tau_r$ = 226.1 ms and a decay time $\tau_d$ = 177.3 ms.

## 3. Conclusions

In summary, we have proposed and experimentally demonstrated a highly responsive infrared photodetector based on multilayer BP, enabled by all-dielectric quasi-BICs, that overcomes intrinsic optical absorption limitations. Structural asymmetry introduced via an arc-shaped notch in the silicon nanopillars excites quasi-BIC resonances, resulting in an ~8-fold enhancement of BP responsivity under 1550 nm illumination relative to the pristine material. The integrated metasurface preserves the intrinsic polarization anisotropy of BP, confirming that the quasi-BIC-enhanced field interacts with the material without altering its native anisotropic response. Leveraging the polarization-dependent response of BP together with the resonant local field enhancement provided by the metasurface enables the design of high-performance photodetectors with enhanced polarization sensitivity. The quasi-BIC-mediated field confinement directly enhances BP photoresponsivity, underscoring the application potential of this architecture for infrared optoelectronics. Integrating symmetry-broken all-dielectric metasurfaces with low-dimensional materials provides an effective route to overcome intrinsic absorption limits, enabling advanced multifunctional infrared photodetectors.

## 4. Materials and methods

### 4.1. Simulation

The optical properties of the plasmonic metasurface were systematically investigated using COMSOL simulations. To ensure consistency with experimental conditions, the simulation geometry was designed to precisely match the structural parameters of the fabricated device. Three frequency-domain monitors were placed to extract key optical responses, including reflection spectra, transmission profiles, and far-field electric field distributions. The simulation setup included periodic boundary conditions along the lateral directions and PML in the vertical direction. The dielectric function of Au and $SiO_2$ were obtained from experimentally validated sources to ensure simulation accuracy.

**4.2. Device Fabrication**

The device fabrication process of the photodetector includes six steps in detail: (a) photoetching and etching to prepare the metasurface, (b) expanding the etching area, (c) preparing the electrode, (d) mechanically peeling to obtain the two-dimensional material, (e) transferring the material to the metasurface, and (f) finally packaging with h-BN, as depicted in **Figure S1**. (a) The metasurface was fabricated on a SOI substrate consisting of a 500 nm top silicon layer and 2 μm buried $SiO_2$ layer. The top Si layer was patterned to produce a metasurface using EBL, in which PMMA A4 photoresist was first spin-coated onto the SOI substrate at 4000 rpm for 30 seconds, followed by ICP etching to define the designed metasurface structures, while the photoresist PMMA A4 acts as a mask during the etching. The equipment used for EBL is JBX5500ZA electron beam mask aligner manufactured by JEOL Company, and the equipment used for ICP is a deep silicon etcher with model STS MUC21 manufactured by SPTS Company. (b) The etched Si region was extended to reserve space for the subsequent deposition using LDWL with AZ5214 photoresist at 4000 rpm for 30 seconds, followed by ICP etching. The device for LDWL is fabricated via an Ultraviolet Maskless Lithography machine (TuoTuo Technology, UV Litho-ACA). (c) Cr/Au electrodes with a thickness of 20 nm/500 nm were deposited using electron beam evaporation. The thickness of the electrode was chosen to be large in order to prevent BP multilayer breaking due to the great difference in surface height between the electrode and the metasurface in the process of material transferring. (d) The multilayer BP sample was prepared by mechanically exfoliating flakes from a crystal onto the adhesive tape. Subsequently, a dry transfer technique was employed to transfer the exfoliated BP onto a Si substrate. During the optical microscopic observation, we were able to select the flakes of appropriate thickness for our experiments. The h-BN flake was also such a process. (e) The PDMS and PC stamp were employed for the deterministic transfer: h-BN was first picked up, followed by BP alignment and pickup at the desired position. The stacked h-BN/BP was then aligned and stamped onto the metasurface under controlled heating conditions to soften the PC layer, enabling intimate contact between the laminated material and the substrate. The adhesion between PMDS and PC film is reduced by heating, and only the PMDS is lifted, leaving the PC film and h-BN/BP material in the metasurface area on the device. All material handling and transfer steps were performed in a nitrogen-filled glovebox to minimize ambient oxygen exposure and prevent oxidation of BP. (f) After transferring, the PC film was dissolved in chloroform, and the device was subsequently annealed in a tube furnace at 180 °C for 3 hours to improve interfacial cleanliness and structural stability, so that the h-BN film played the role of encapsulation to protect the BP layer in the covered area from oxidation.

## 4.3. Characterizations

The AFM curve was performed by Atomic Force Microscope (Dimension ICON, Bruker) using Tapping mode. Raman spectroscopy was conducted by Raman Spectrometer (LabRAM HR Evolution, Horiba Jobin Yvon) with 532 nm laser excitation. Electrical characterizations were acquired by a semiconductor characterization system (4200A-SCS, Keithley). The laser output is focused into a compact spot via the convex lens, redirected by reflectors, filtered through the polarizer, and directed onto the sample; precise localization of the irradiation spot is achieved using the microscope to target regions with/without the metasurface on BP. The optical power is measured by a laser power meter (PM400k3, THORLABS). This configuration enables accurate manipulation of laser conditions and sample alignment, indicating its suitability for dependable variable isolation and assessment of metasurface-induced enhancements. All electrical measurements were carried out at room temperature under ambient conditions. The SEM images were measured by using an emission scanning electron microscope (Nova NanoSEM450, Thermo Fisher Scientific) of the Vacuum Interconnected Nanotech Workstation. The femtosecond transient absorption spectroscopy measurements were performed using a setup comprising a regeneratively amplified Ti: sapphire laser source (central wavelength: 800 nm, pulse duration: 35 fs, pulse energy: 6 mJ, repetition rate: 1 kHz), nonlinear frequency mixing for wavelength tuning, and a spectrometer (Helios, Ultrafast Systems) for detection. All TA-spectra measurements were carried out at room temperature under ambient conditions. The angle-resolved spectroscopy was performed by Angle-Resolved Micro-Spectrometer and Optical Measurement System (ARMS and Metronles-F, Shanghai Ideaoptics Corp., Ltd.) of Melalens of Instrumentation and Service Center for Molecular Sciences in Westlake University.

## ASSOCIATED CONTENT

**Supporting Information**

Experimental section and supporting data including device preparation process, SEM observation, angle-resolved spectroscopy of different notched depth, dark current and femtosecond transient absorption (TA) spectroscopy (PDF).

## AUTHOR INFORMATION

**Corresponding Authors**

*Junjia Wang – National Research Center for Optical Sensing/Communications Integrated Networking, School of Electronic Science and Engineering, Southeast University, Nanjing 210096, P. R. China; orcid.org/0000-0003-0872-4794; Email: junjia_wang@seu.edu.cn;*

*Xuechao Yu – Key Laboratory of Multifunctional Nanomaterials and Smart Systems, Suzhou Institute of Nano-Tech and Nano-Bionics, Chinese Academy of Sciences, Suzhou 215123, P. R. China; orcid.org/0000-0003-1425-1911; Email: xcyu2022@sinano.ac.cn;*

*Hongliang Li – Key Laboratory of Multifunctional Nanomaterials and Smart Systems, Suzhou Institute of Nano-Tech and Nano-Bionics, Chinese Academy of Sciences, Suzhou 215123, P. R. China; Email: hlli2025@sinano.ac.cn;*

**Authors**

*Xiao Liu – School of integrated circuits, Southeast University, Wuxi 214026, P. R. China; Email: 236300@sinano.ac.cn;*

*Tianxiang Zhao – School of electronic science & engineering, Southeast University, Nanjing 210096, P. R. China; Email: zhaotx@seu.edu.cn;*

*Ting Wang – Nano Science and Technology Institute, University of Science and Technology of China, Hefei 230026, P. R. China; Email: twang2022@sinano.ac.cn;*

*Junsheng Xu – Nano Science and Technology Institute, University of Science and Technology of China, Hefei 230026, P. R. China; Email: jsxu2024@sinano.ac.cn;*

*Junyong Wang – i-Lab, Suzhou Institute of Nano-Tech and Nano-Bionics, Chinese Academy of Sciences, Suzhou 215123, P. R. China; Email: jywang2022@sinano.ac.cn;*

*Kai Zhang – i-Lab, Suzhou Institute of Nano-Tech and Nano-Bionics, Chinese Academy of Sciences, Suzhou 215123, P. R. China; Email: kzhang2015@sinano.ac.cn;*

**Author Contributions**

The ideas and supervision for this article were contributed by X.Y., J.W.* and H.L. X.L. and T.Z. performed the characterization, investigation, visualization, editing and writing. T.W. and J.X. performed the material transferring under the guidance of K.Z. and J.W. All authors have approved the final version of the manuscript. J.W.* refers to Junjia Wang and J.W. refers to Junyong Wang.

**Funding Sources**

This work was financially supported by the National Key R&D Program of China (Grant No. 2025YFF0524700), the National Natural Science Foundation of China (Grant No. 62375051, 62505351, 62375279), the Basic Research Program of Jiangsu (Grant No. BK20250501, BK20240125), the Jiangsu Funding Program for Excellent Postdoctoral Talent (Grant No. 2025ZB653), and the Suzhou Basic Research Program (Grant No. SYG202340, SJC2023004, ZXL2324340).

**Notes**

The authors declare no competing financial interest.

**ACKNOWLEDGEMENTS**

The authors are grateful for the technical support from the Nano-X from Suzhou Institute of Nano-Tech and Nano-Bionics, Chinese Academy of Sciences (SINANO). The devices were characterized by ATS-PC-03 (TuoTuo Technology, ATS-PC-03).

**ABBREVIATIONS**

FDTD, finite-difference time-domain; SOI, silicon on insulator; ICP, inductively coupled plasma; AFM, atomic force microscopy; SEM, scanning electron microscope;

PML, perfectly matched layer; EBL, electron beam lithography; LDWL, laser direct write lithography; PDMS, polydimethylsiloxane; PC, polycarbonate.

**REFERENCES**

(1) Flöry, N.; Ma, P.; Salamin, Y.; Emboras, A.; Taniguchi, T.; Watanabe, K.; Leuthold, J.; Novotny, L. Waveguide-integrated van der Waals heterostructure photodetector at telecom wavelengths with high speed and high responsivity. *Nature Nanotechnology* **2020**, *15*, 118-124.

(2) Ma, Y.; Shan, L.; Ying, Y.; Shen, L.; Fu, Y.; Fei, L.; Lei, Y.; Yue, N.; Zhang, W.; Zhang, H. Day-Night imaging without Infrared Cutfilter removal based on metal-gradient perovskite single crystal photodetector. *Nat. Commun.* **2024**, *15*, 7516.

(3) Ma, C.; Yuan, S.; Cheung, P.; Watanabe, K.; Taniguchi, T.; Zhang, F.; Xia, F. Intelligent infrared sensing enabled by tunable moiré quantum geometry. *Nature* **2022**, *604*, 266-272.

(4) Martyniuk, P.; Wang, P.; Rogalski, A.; Gu, Y.; Jiang, R.; Wang, F.; Hu, W. Infrared avalanche photodiodes from bulk to 2D materials. *Light: Science & Applications* **2023**, *12*, 212.

(5) Xia, F.; Mueller, T.; Lin, Y.-m.; Valdes-Garcia, A.; Avouris, P. Ultrafast graphene photodetector. *Nature nanotechnology* **2009**, *4*, 839-843.

(6) Liu, H.; Du, Y.; Deng, Y.; Ye, P. D. Semiconducting black phosphorus: synthesis, transport properties and electronic applications. *Chem. Soc. Rev.* **2015**, *44*, 2732-2743.

(7) Gao, P.; Liu, S.; Tang, Y.; Wang, T.; Ran, S.; Liu, L.; Zhang, M.; Tan, F.; Chen, J.; Wei, J. High-Performance Polarization-Sensitive Photodetectors Based on Fully

Depleted T‑MoS$_2$/MoTe$_2$/B‑MoS$_2$ Heterojunction. *Advanced Functional Materials* **2025**, e14834.

(8) Lim, K. R. G.; Shekhirev, M.; Wyatt, B. C.; Anasori, B.; Gogotsi, Y.; Seh, Z. W. Fundamentals of MXene synthesis. *Nature Synthesis* **2022**, *1*, 601-614.

(9) Zhu, X.; Cai, Z.; Wu, Q.; Wu, J.; Liu, S.; Chen, X.; Zhao, Q. 2D Black Phosphorus Infrared Photodetectors. *Laser & Photonics Reviews* **2025**, *19*, 2400703.

(10) Wang, X.; Lan, S. Optical properties of black phosphorus. *Advances in Optics and Photonics* **2016**, *8*, 618-655.

(11) Mu, X.; Wang, J.; Sun, M. Two-dimensional black phosphorus: physical properties and applications. *Materials Today Physics* **2019**, *8*, 92-111.

(12) Mao, N.; Tang, J.; Xie, L.; Wu, J.; Han, B.; Lin, J.; Deng, S.; Ji, W.; Xu, H.; Liu, K.; Tong, L.; Zhang, J. Optical Anisotropy of Black Phosphorus in the Visible Regime. *J. Am. Chem. Soc.* **2016**, *138*, 300-305.

(13) Amani, M.; Regan, E.; Bullock, J.; Ahn, G. H.; Javey, A. Mid-Wave Infrared Photoconductors Based on Black Phosphorus-Arsenic Alloys. *ACS Nano* **2017**, *11*, 11724-11731.

(14) Guo, Q.; Pospischil, A.; Bhuiyan, M.; Jiang, H.; Tian, H.; Farmer, D.; Deng, B.; Li, C.; Han, S.-J.; Wang, H. Black phosphorus mid-infrared photodetectors with high gain. *Nano letters* **2016**, *16*, 4648-4655.

(15) Chen, X.; Lu, X.; Deng, B.; Sinai, O.; Shao, Y.; Li, C.; Yuan, S.; Tran, V.; Watanabe, K.; Taniguchi, T. Widely tunable black phosphorus mid-infrared photodetector. *Nat. Commun.* **2017**, *8*, 1672.

(16) Xu, Y.; Liu, C.; Guo, C.; Yu, Q.; Guo, W.; Lu, W.; Chen, X.; Wang, L.; Zhang, K. High performance near infrared photodetector based on in-plane black phosphorus pn homojunction. *Nano Energy* **2020**, *70*, 104518.

(17) Yin, Y.; Cao, R.; Guo, J.; Liu, C.; Li, J.; Feng, X.; Wang, H.; Du, W.; Qadir, A.; Zhang, H. High‐speed and high‐responsivity hybrid silicon/black‐phosphorus waveguide photodetectors at 2 μm. *Laser & Photonics Reviews* **2019**, *13*, 1900032.

(18) Chang, T.-Y.; Chen, P.-L.; Yan, J.-H.; Li, W.-Q.; Zhang, Y.-Y.; Luo, D.-I.; Li, J.-X.; Huang, K.-P.; Liu, C.-H. Ultra-broadband, high speed, and high-quantum-efficiency photodetectors based on black phosphorus. *ACS Appl. Mater. Interfaces* **2019**, *12*, 1201-1209.

(19) Li, X.; Yu, Z.; Xiong, X.; Li, T.; Gao, T.; Wang, R.; Huang, R.; Wu, Y. High-speed black phosphorus field-effect transistors approaching ballistic limit. *Science advances* **2019**, *5*, eaau3194.

(20) Wang, L.; Liu, C.; Chen, X.; Zhou, J.; Hu, W.; Wang, X.; Li, J.; Tang, W.; Yu, A.; Wang, S. W. Toward sensitive room‐temperature broadband detection from infrared to terahertz with antenna‐integrated black phosphorus photoconductor. *Advanced Functional Materials* **2017**, *27*, 1604414.

(21) Zhao, X.; Lou, T.; Peng, Y.; Sun, F.; Wei, X. Metasurface-Based Photodetectors: Pursuing Superior Performance and Multifunctionality. *ACS Photonics* **2025**, *12*, 4096-4118.

(22) Fang, J.; Wang, D.; DeVault, C. T.; Chung, T.-F.; Chen, Y. P.; Boltasseva, A.; Shalaev, V. M.; Kildishev, A. V. Enhanced graphene photodetector with fractal metasurface. *Nano letters* **2017**, *17*, 57-62.

(23) Jiang, H.; Chen, Y.; Guo, W.; Zhang, Y.; Zhou, R.; Gu, M.; Zhong, F.; Ni, Z.; Lu, J.; Qiu, C.-W.; Gao, W. Metasurface-enabled broadband multidimensional photodetectors. *Nat. Commun.* **2024**, *15*, 8347.

(24) Zhao, H.; Wang, W.; Ding, H.; Wang, S.; Tang, Z.; Li, S.; Wang, J.; Wang, Y.; Zhou, Q.; Wang, A.; Yu, Y.; Gao, L. Dielectric metasurface enhanced performance in multilayer $WS_2$ photodetector. *Science China Information Sciences* **2025**, *68*, 179403.

(25) Yu, N.; Capasso, F. Flat optics with designer metasurfaces. *Nat. Mater.* **2014**, *13*, 139-150.

(26) Chen, H.-T.; Taylor, A. J.; Yu, N. A review of metasurfaces: physics and applications. *Rep. Prog. Phys.* **2016**, *79*, 076401.

(27) Guan, R.; Xu, H.; Lou, Z.; Zhao, Z.; Wang, L. Design and development of metasurface materials for enhancing photodetector properties. *Adv. Sci.* **2024**, *11*, 2402530.

(28) Li, L.; Wang, J.; Kang, L.; Liu, W.; Yu, L.; Zheng, B.; Brongersma, M. L.; Werner, D. H.; Lan, S.; Shi, Y. Monolithic full-Stokes near-infrared polarimetry with chiral plasmonic metasurface integrated graphene-silicon photodetector. *ACS nano* **2020**, *14*, 16634-16642.

(29) Lien, M. R.; Wang, N.; Guadagnini, S.; Wu, J.; Soibel, A.; Gunapala, S. D.; Wang, H.; Povinelli, M. L. Black Phosphorus Molybdenum Disulfide Midwave Infrared Photodiodes with Broadband Absorption-Increasing Metasurfaces. *Nano Letters* **2023**, *23*, 9980-9987.

(30) Huo, P.; Song, M.; Zhu, W.; Zhang, C.; Chen, L.; Lezec, H. J.; Lu, Y.; Agrawal, A.; Xu, T. Photorealistic full-color nanopainting enabled by a low-loss metasurface. *Optica* **2020**, *7*, 1171-1172.

(31) Yang, Y.; Kravchenko, I. I.; Briggs, D. P.; Valentine, J. All-dielectric metasurface analogue of electromagnetically induced transparency. *Nat. Commun.* **2014**, *5*, 5753.

(32) Yu, Y. F.; Zhu, A. Y.; Paniagua‐Domínguez, R.; Fu, Y. H.; Luk'yanchuk, B.; Kuznetsov, A. I. High‐transmission dielectric metasurface with 2π phase control at visible wavelengths. *Laser & Photonics Reviews* **2015**, *9*, 412-418.

(33) Li, P.; Dolado, I.; Alfaro-Mozaz, F. J.; Casanova, F.; Hueso, L. E.; Liu, S.; Edgar, J. H.; Nikitin, A. Y.; Vélez, S.; Hillenbrand, R. Infrared hyperbolic metasurface based on nanostructured van der Waals materials. *Science* **2018**, *359*, 892-896.

(34) King, J.; Wan, C.; Park, T. J.; Deshpande, S.; Zhang, Z.; Ramanathan, S.; Kats, M. A. Electrically tunable $VO_2$-metal metasurface for mid-infrared switching, limiting and nonlinear isolation. *Nature Photonics* **2024**, *18*, 74-80.

(35) Koshelev, K.; Bogdanov, A.; Kivshar, Y. Meta-optics and bound states in the continuum. *Science Bulletin* **2019**, *64*, 836-842.

(36) Zhou, Q.; Fu, Y.; Huang, L.; Wu, Q.; Miroshnichenko, A.; Gao, L.; Xu, Y. Geometry symmetry-free and higher-order optical bound states in the continuum. *Nat. Commun.* **2021**, *12*, 4390.

(37) Fang, Y. t.; Bu, F.; He, S. Abnormal Unidirectional Lasing from the Combined Effect of non-Hermitian Modulated Bound States in the Continuum and Fabry-Pérot Resonance. *Laser & Photonics Reviews* **2025**, *19*, 2400964.

(38) Doiron, C. F.; Brener, I.; Cerjan, A. Realizing symmetry-guaranteed pairs of bound states in the continuum in metasurfaces. *Nat. Commun.* **2022**, *13*, 7534.

(39) Zhou, C.; Huang, L.; Jin, R.; Xu, L.; Li, G.; Rahmani, M.; Chen, X.; Lu, W.; Miroshnichenko, A. E. Bound states in the continuum in asymmetric dielectric metasurfaces. *Laser & Photonics Reviews* **2023**, *17*, 2200564.

(40) Wang, T.; Di, W.; Sha, W. E. I.; Proietti Zaccaria, R. Enabling low Threshold Laser Through an Asymmetric Tetramer Metasurface Harnessing Polarization-Independent Quasi-BICs. *Advanced Optical Materials* **2025**, *13*, 2403345.

(41) Xu, L.; Zangeneh Kamali, K.; Huang, L.; Rahmani, M.; Smirnov, A.; Camacho-Morales, R.; Ma, Y.; Zhang, G.; Woolley, M.; Neshev, D.; Miroshnichenko, A. E. Dynamic Nonlinear Image Tuning through Magnetic Dipole Quasi-BIC Ultrathin Resonators. *Adv. Sci.* **2019**, *6*, 1802119.

(42) Sun, K.; Jiang, H.; Bykov, D. A.; Van, V.; Levy, U.; Cai, Y.; Han, Z. 1D quasi-bound states in the continuum with large operation bandwidth in the ω~k space for nonlinear optical applications. *Photonics Research* **2022**, *10*, 1575.

(43) Tan, T. C.; Srivastava, Y. K.; Ako, R. T.; Wang, W.; Bhaskaran, M.; Sriram, S.; Al-Naib, I.; Plum, E.; Singh, R. Active Control of Nanodielectric-Induced THz Quasi-BIC in Flexible Metasurfaces: A Platform for Modulation and Sensing. *Adv. Mater.* **2021**, *33*, 2100836.

(44) Wu, K.; Zhang, C.; Yu, T.; Luo, Y.; Wu, L.; Li, S.; Chen, Z.; Zhu, X.; Yang, J.; Li, X. Highly Polarized Near-Infrared Photodetection in Monolayer $WSe_2$ Enabled by Plasmonic Bound States in the Continuum. *ACS Photonics* **2025**, *12*, 6226-6236.

(45) Wang, Y.; Yu, Z.; Zhang, Z.; Sun, B.; Tong, Y.; Xu, J.-B.; Sun, X.; Tsang, H. K. Bound-States-in-Continuum Hybrid Integration of 2D Platinum Diselenide on Silicon Nitride for High-Speed Photodetectors. *ACS Photonics* **2020**, *7*, 2643-2649.

(46) Bullock, J.; Amani, M.; Cho, J.; Chen, Y.-Z.; Ahn, G. H.; Adinolfi, V.; Shrestha, V. R.; Gao, Y.; Crozier, K. B.; Chueh, Y.-L. Polarization-resolved black phosphorus/molybdenum disulfide mid-wave infrared photodiodes with high detectivity at room temperature. *Nature Photonics* **2018**, *12*, 601-607.

(47) Guo, J.; Jin, R.; Fu, Z.; Zhang, Y.; Yu, F.; Chen, J.; Wang, X.; Huang, L.; Zhou, C.; Chen, X. Topologically engineered high-Q quasi-BIC metasurfaces for enhanced near-infrared emission in PbS quantum dots. *Nano Letters* **2025**, *25*, 2357-2365.

(48) Hu, C.; Liu, X.; Liu, G.; Wang, X.; Li, C.; Chen, J.; Tang, C.; Deng, J.; Liu, Z. High-Q electromagnetically induced transparency and slow-light manipulation via utilizing the quasi-BICs. *Opt. Commun.* **2025**, 132545.

(49) Zeng, D.; Wu, Z.; Liu, G.; Yu, M.; Liu, X.; Chen, J.; Tang, C.; Du, W.; Liu, Z. Ultra-slow-light and dynamically quantitative optical storage modulation via quasi-BICs. *Opt. Lett.* **2024**, *49*, 3030-3033.

(50) McMahon, J. M.; Kioupakis, E.; Schulz, S. Atomistic analysis of Auger recombination in *c*-plane (In,Ga)N/GaN quantum wells: Temperature-dependent competition between radiative and nonradiative recombination. *Physical Review B* **2022**, *105*, 195307.

(51) Iveland, J.; Martinelli, L.; Peretti, J.; Speck, J. S.; Weisbuch, C. Direct Measurement of Auger Electrons Emitted from a Semiconductor Light-Emitting Diode under Electrical Injection: Identification of the Dominant Mechanism for Efficiency Droop. *Phys. Rev. Lett.* **2013**, *110*, 177406.

(52) Klimov, V. I.; Mikhailovsky, A. A.; McBranch, D.; Leatherdale, C. A.; Bawendi, M. G. Quantization of multiparticle Auger rates in semiconductor quantum dots. *Science* **2000**, *287*, 1011-1013.

(53) Delaney, K. T.; Rinke, P.; Van de Walle, C. G. Auger recombination rates in nitrides from first principles. *Appl. Phys. Lett.* **2009**, *94*.

(54) Holland, M. G. Analysis of Lattice Thermal Conductivity. *Phys. Rev.* **1963**, *132*, 2461-2471.

(55) Che, Y.; Zhang, T.; Liu, X.; Hu, D.; Song, S.; Cai, Y.; Cao, Y.; Zhang, J.; Chu, S.-W.; Li, X. Nanophotonic inspection of deep-subwavelength integrated optoelectronic chips. *Science Advances* **2025**, *11*, eadr8427.

(56) M, P.; Pandey, S.; R, K.; Sati, A.; Trivedi, R.; Raviprakash, Y.; Kamath, S. D.; Mishra, V. Uncovering temperature-induced changes in bandgap and electronic heterogeneity in transition metal oxides through optical absorption spectroscopy: A review. *Physica B: Condensed Matter* **2024**, *695*, 416485.

(57) Ščajev, P.; Miasojedovas, S.; Subačius, L.; Jarašiūnas, K.; Mazanik, A. V.; Korolik, O. V.; Kato, M. Impact of intrinsic defects on excitation dependent carrier lifetime in thick 4H-SiC studied by complementing microwave photoconductivity, free-carrier absorption and time-resolved photoluminescence techniques. *J. Lumin.* **2019**, *212*, 92-98.

(58) Wu, D.; Xu, M.; Zeng, L.; Shi, Z.; Tian, Y.; Li, X. J.; Shan, C.-X.; Jie, J. In situ fabrication of $PdSe_2$/GaN Schottky junction for polarization-sensitive ultraviolet photodetection with high dichroic ratio. *ACS nano* **2022**, *16*, 5545-5555.

(59) Xin, W.; Zhong, W.; Shi, Y.; Shi, Y.; Jing, J.; Xu, T.; Guo, J.; Liu, W.; Li, Y.; Liang, Z. Low‐dimensional‐materials‐based photodetectors for next‐generation polarized detection and imaging. *Adv. Mater.* **2024**, *36*, 2306772.

(60) Wang, J.; Jiang, C.; Li, W.; Xiao, X. Anisotropic low‐dimensional materials for polarization-sensitive photodetectors: from materials to devices. *Advanced Optical Materials* **2022**, *10*, 2102436.

(61) Seah, S. K.; Biswas, S.; Hail, C. U.; Jang, M. S.; Atwater, H. A. Efficient, Polarization-Diverse State Generation with Tunable Black Phosphorus Cavity Heterostructures and Active Metasurfaces. *ACS Photonics* **2025**, *12*, 1409-1417.

(62) Yuan, H.; Liu, X.; Afshinmanesh, F.; Li, W.; Xu, G.; Sun, J.; Lian, B.; Curto, A. G.; Ye, G.; Hikita, Y. Polarization-sensitive broadband photodetector using a black phosphorus vertical p-n junction. *Nature nanotechnology* **2015**, *10*, 707-713.

(63) Biswas, S.; Grajower, M. Y.; Watanabe, K.; Taniguchi, T.; Atwater, H. A. Broadband electro-optic polarization conversion with atomically thin black phosphorus. *Science* **2021**, *374*, 448-453.

(64) Kim, J.; Lee, J.-U.; Lee, J.; Park, H. J.; Lee, Z.; Lee, C.; Cheong, H. Anomalous polarization dependence of Raman scattering and crystallographic orientation of black phosphorus. *Nanoscale* **2015**, *7*, 18708-18715.